\documentclass[superscriptaddress,aps,preprint,amsmath,amssymb,floatfix]{revtex4}
\usepackage{graphicx,morefloats}
\usepackage{subfigure} 
\usepackage{xcolor,soul}
\usepackage{caption}
\usepackage{bm}
\usepackage{helvet}
\usepackage{hyperref}

\definecolor{bg0}{rgb}{0,0,0}
\definecolor{bg}{rgb}{0,0,0.3}
\definecolor{bg1}{rgb}{1,1,1}
\definecolor{darkgreen}{rgb}{0,0.6,0.15}
\definecolor{mycolor}{rgb}{0.98, 0.0, 0.75}

\usepackage{caption,newfloat}
\DeclareCaptionType{InfoBox}

\begin{document}


\hyphenation{va-ni-sh-ing ana-ly-zing}

\begin{center}

\thispagestyle{empty}

{\large\bf Iron-based superconductors: teenage, complex, challenging
}
\\[0.3cm]

Qimiao Si$^{1}$ and  Nigel E. Hussey$^{2,3}$\\[0.3cm]

$^1$Department of Physics and Astronomy, Rice Center for Quantum Materials, \\
Rice University, Houston, TX 77005,
USA\\

$^2$H. H. Wills Physics Laboratory, University of Bristol, Tyndall Avenue, Bristol BS8 1TL, United Kingdom\\

$^3$High Field Magnet Laboratory (HFML-EMFL) and Institute for Molecules and Materials, Radboud University, Toernooiveld 7, 6525 ED Nijmegen, Netherlands\\

\end{center}

\vspace{0.0cm}

{\bf  The advent of iron-based superconductors in 2008 came as a complete surprise to the 
condensed matter community. Now  15 years later, they are 
beginning to impart some of their new-found wisdom on a slew 
of emerging superconductors that boast similar traits.
}

\vspace{0.4cm}

\noindent E-mails: qmsi@rice.edu; n.e.hussey@bristol.ac.uk


\section{Introduction}\label{introduction}

The Annual Meeting of the American Physical Society in March 2008 was especially noteworthy
for condensed matter physicists. It 
seemed that all that anyone wanted to talk about at that
meeting in New Orleans was the surprising discovery, and subsequent confirmation, of high 
temperature superconductivity in the iron arsenides\cite{hosono}. 
Surprising
was perhaps something of an 
understatement. After all, iron was supposed to be as toxic to superconductivity as arsenic 
was to humans. The superconducting transition temperature $T_c$ had reached only 26 K by the 
time of the 
March
meeting though reports were circulating that, as with high-$T_c$ cuprates two 
decades earlier, the application of high pressures was pushing $T_c$ to temperatures above 
40 K. It was clear the race was on and in a matter of weeks after the 
meeting,
 substitutional 
studies had led to the $T_c$ record being broken repeatedly, eventually plateauing at an 
eye-watering 56 K. As a barometer of the excitement generated by this new research line, 
by the end of 2010, 8 out of 10 of the most cited papers published in {\it Physical Review 
Letters} in 2008 featured 
arsenide
 superconductivity. Amidst all this frenzied
activity,
superconductivity was also being reported in a simple binary 
iron chalcogenide FeSe. While the $T_c$ of FeSe was relatively modest -- at only 8 K in 
the bulk -- in monolayer form, it went on to claim the record $T_c$ of all the 
iron-based 
superconductors (FeSCs) with a $T_c$ in excess of 65 K. To date, only the cuprates are known 
to possess a higher $T_c$ at ambient pressure and while the maximum $T_c$ in either system 
has not shifted in the last decade, the field of superconductivity is currently enjoying a 
resurgence, spurred on by the discovery of unconventional and/or 
high-temperature superconductivity in a host of new materials and extreme environments.

In light of this, it
feels
 an opportune moment to provide a status update 
 on
  the 
 FeSCs.
Because the early years of FeSC research were expertly surveyed 
in {\it Physics Today} back in
August 2009 (by Charles Day) and again in 2015 (by Andrey Chubukov and Peter Hirschfeld),
our focus here will be on developments that have come to light in the intervening years. During 
this period, the
 all-encompassing effect
 of electron correlations has been recognized,
thereby greatly enriching the field.
As is typical for systems where correlation effects are important, 
puzzles and surprises abound.
Indeed, it
has 
 become abundantly clear that the FeSCs provide a setting that, while challenging, allows 
for a deeper understanding of such diverse themes as electronic nematicity
(i.e., orientational order), quantum 
criticality, orbital-selective correlations
and topological 
superconductivity (we 
will elaborate on these terms later in the article).
Thus, even though FeSCs are 
still in their teenage 
years, the maturity of the field is such that they are already providing profound insights into 
the physics of unconventional superconductivity, physics that may well be playing a key role in 
other
emerging superconductors.


\section{Some basics}\label{sec:basics}

Iron-based superconductors primarily comprise iron pnictides, compounds based on arsenic or another element from the pnictogen group, and iron chalcogenides containing selenium, tellurium or sulfur. There
 is a broad spectrum of structural types across which $T_c$ has a 
large spread (Box 1)\cite{basics}.  
The simplest structure occurs in FeSe, in which individual FeSe layers
stack on top of each other. 

It is well-established that 
the highest occupied
electronic states 
 are
 derived almost entirely from the 5-fold 
 $3d$
 orbital
  states of the Fe ions. The existence of Se or 
As means that each unit cell 
 comprises two Fe ions. 
When spin-orbit coupling (which is relatively small for $3d$-electron-based systems)
is neglected, the crystalline symmetry allows us to consider a simpler unit cell containing only 
one Fe ion per unit cell. These Fe ions form a square lattice, which is relatively straightforward 
to handle theoretically. 

In any crystal, atoms form a periodic arrangement leading to a crystalline lattice. 
The discrete lattice in real space
implies, by virtue of a Fourier transform, a corresponding lattice in reciprocal, 
i.e. wavevector, space. Condensed 
matter physicists call the unit cell of this reciprocal space a Brillouin zone. 
Only a single Brillouin zone is needed
to account for all the electronic states of 
the crystal. For a metal, the Pauli principle
dictates that each wavevector is associated with 
an individual and distinct set of internal
({\it e.g.}, spin and orbital)
 quantum numbers.
  For a noninteracting 
electron system, the electrons occupy the states with wavevectors that 
are associated with an increasing ladder of 
energies. The locus of wavevectors corresponding to the highest energy of the occupied states 
(known as the Fermi energy) forms a Fermi surface.
The Brillouin zone of a square lattice is another
simple square. 
For a number of 
iron chalcogenides in the group of highest-$T_c$ FeSCs, including monolayer FeSe, the Fermi surface 
also
turns out to be remarkably simple, comprising only small electron pockets 
located
 the 
edge of the Brillouin zone
(see Box 1, panel D).


\section{Electron correlations}\label{sec:el_cor}

When two electrons occupy either the same $3d$ orbital or two different $3d$ orbitals of an Fe ion, 
their close proximity inevitably leads to a Coulomb repulsive interaction between them. The size of 
this 
 repulsion quantifies the strength of electron correlations in the system
 which in turn can cause the electrons to become heavier and slower.
 It was recognized 
early on that electron correlations are important to the physics of FeSCs\cite{basics}.
One manifestation of strong correlations is an electrical resistivity that exhibits 
so-called `bad metal'
behavior characterized by an anomalously short electron mean-free-path at room temperature \cite{hussey04}, 
as described in Fig.\,1.
 Accompanying evidence for 
 this
 comes from measurements of the optical conductivity
and
angle-resolved photoemission spectroscopy (ARPES), which 
see
 a significant
 renormalization
 of, among other parameters, 
 the effective electron mass $m^\ast$ relative to its noninteracting counterpart $m_b$\cite{basics}.
 Remarkably, in FeSCs, this renormalization can even be orbital specific with
 the mass enhancement $m^{\ast}/m_b$ 
 varying
  strongly
  from one orbital to another.
In the 
Fe(Te,Se)
series, $m^{\ast}/m_b$ 
in one of its $3d$ orbitals
reaches 
almost 
ten times that for the
other
 $3d$ orbitals\cite{yi-osm}. 

Theoretical treatments of the electron correlations in 
 models that contain multiple $3d$ orbitals
anticipated this 
orbital selectivity or differentiation.
One theoretical approach considers the system to be in
proximity to an orbital-selective 
Mott 
 phase 
where electrons within specific orbitals are localized on their respective lattice 
sites, 
while the others remain itinerant,
depending 
on the relative strength of the (unscreened) Coulomb repulsion
(see Fig.\,2A \cite{theory-osm}).
Another approach
 proposes a
so-called Hund's metal\cite{os2},
with localized spin excitations but fully itinerant charge and orbital excitations.
Because the dominant 
Coulomb repulsion is local
({\it i.e.}, not long-ranged),
 the
  electron dynamics 
  are best
 captured
  in terms of short-range hopping between adjacent atomic sites rather than the usual metallic, wave-like propagation across the whole sample.
Excitations between
 atomic
energy levels
split by the Coulomb repulsion create an additional peak in the excitation spectrum. 
Observations of this additional peak
 \cite{watson-lhb,yi-osm},
 illustrated in Fig.\,2B,
also 
provide direct evidence that the local Coulomb repulsion is 
large.
At the same time, 
  the propensity for
   Fermi-surface nesting (Box 1), 
 whereby 
 extended
  sections of the Fermi surface are linked through a single wavevector,
 also enhances correlations effects by virtue of influencing a large fraction of the electronic states\cite{nesting}.


\section{Electronic order and quantum criticality}\label{sec:order_qc}

Classical phase transitions in matter take place when temperature is changed, 
as exemplified by ice melting or water evaporating. Quantum phase transitions, by contrast, 
 occur
 at zero temperature, 
and are induced by the change in the degree to which Heisenberg's uncertainty principle is 
manifested upon the variation of a non-thermal control parameter. Quantum criticality develops when 
the transition is continuous, and controls the physics in a larger parameter regime at nonzero temperatures.
In the iron pnictides, superconductivity adjoins 
an antiferromagnetically
 ordered phase
 in which the adjacent spins are anti-aligned along a specific direction\cite{basics}. By tuning 
each system appropriately, either through chemical substitution or by applying pressure, the phase transition 
to the magnetic state at 
the N\'{e}el temperature
 is suppressed. Though superconductivity invariably intervenes,
experimental signatures of mass renormalization in its vicinity imply
that the magnetic phase terminates at a 
quantum
 critical point (QCP), where quantum fluctuations 
destroy the magnetic order, even at zero temperature (Box 2)\cite{qcp-FeAs}. 

   Earlier transport and thermodynamic 
measurements on isovalently-substituted BaFe$_2$(As$_{1-x}$P$_x$)$_2$ provided strong evidence for mass 
renormalization due to the interaction of the itinerant carriers with quantum critical fluctuations of the 
underlying order parameter\cite{qcp-FeAs}. 
These measurements could not, however, establish the character 
of the fluctuating order. In addition to magnetic order, iron pnictides also possess 
nematic
order that
spontaneously breaks the symmetry between the $x$ and $y$ spatial directions,
 setting in at or around the tetragonal-to-orthorhombic structural 
transition temperature $T_S$. We illustrate the emergence of nematic order in FeSe in Fig.\,3. 
The nematic fluctuations occur over a wide energy range (about 50 meV, which is sizable compared 
to the $\sim 200$-meV scale of magnetic fluctuations), 
underscoring
 their 
importance
\cite{qcp-FeAs-nematic}.
 From a theoretical perspective, nematic fluctuations have been shown 
to enhance any pairing interaction that is present and
thus
could play a key role in increasing the critical 
temperature even if the pairing
itself
 is primarily driven by quantum fluctuations in other, e.g. spin, 
channels\cite{nematicity-driving-sc}.

In many cases, the nematicity and magnetism appear to be strongly intertwined (Box 2). 
More recently, however, 
researchers have gone in search of signatures of pure nematic quantum criticality. 
In non-magnetic FeSe, 
  nematic order 
  emerges 
at $T_S$, 
 below which both its normal and 
superconducting state properties exhibit marked two-fold anisotropy.
$T_S$ can be suppressed either through the application of high pressures, or via chemical substitution on 
the chalcogenide site. With increasing pressure, the critical nematic fluctuations in FeSe are quenched, 
presumably due to the emergence of long-range magnetic order before the nematic phase 
terminates. 
 As mentioned above
$T_c$ grows more than four-fold up to 40 K
with increasing pressure,
 in tandem with the strengthening magnetic interactions. In 
sulfur-substituted FeSe, nematicity vanishes at a critical S concentration $x_c$ 
at which point
 the nematic susceptibility as deduced from elasto-resistivity measurements
also diverges\cite{qcp-FeSe}. Since no magnetic order develops at any point across the substitution series 
(at ambient pressure), this divergence hints at a 
pure
nematic QCP in FeSe$_{1-x}$S$_x$. 
One of the key experimental signatures of quantum criticality is an electrical resistivity displaying a marked 
linear temperature dependence down to low temperatures (well below the typical temperature scales associated 
with electron-phonon scattering),
an example of which is
shown in Fig.\,1. 
Such behavior 
may also
be linked with the bad-metal transport seen at high temperatures.
The 
$T$-linear resistivity
 seen at ambient pressure in FeSe$_{1-x}$S$_x$ has thus been attributed to the emergent critical nematic fluctuations, though residual spin fluctuations (believed to be responsible for the $T$-linear resistivity seen in BaFe$_2$(As$_{1-x}$P$_x$)$_2$) have not been
 completely ruled out.

In addition to this
 $T$-linear resistivity, magnetotransport studies on iron-based superconductors 
have uncovered a startling new feature of
quantum critical transport,
namely
 a crossover to 
linear-in-field magnetoresistance in systems close to 
the QCP\cite{Hayes}.
 The particular scaling
 form of 
the magnetoresistance
(see inset to Fig.\,1)
suggests
 an intimate, but as yet unresolved, connection to the 
 $T$-linear resistivity at zero field. 
 

\section{Unconventional superconductivity}
\label{sec:sc}

 In the iron pnictides, the close proximity of the superconducting phase to the 
static magnetic order suggests that the quantum fluctuations associated with the antiferromagnetic spin-exchange 
interactions play an important role in the pairing mechanism. The same may apply to the iron chalcogenides: 
In the part of the phase diagram where superconductivity develops, the nature of the electronic order differs 
from that seen in the iron pnictides, yet antiferromagnetic fluctuations are still prevalent. 
 While smoking-gun evidence for 
nematic-fluctuation-assisted superconductivity remains elusive, equally there is little evidence out there to 
suggest that nematic quantum criticality is bad for superconductivity. On the contrary, superconductivity often 
emerges in 
FeSCs
 that exhibits quantum critical nematic fluctuations.

In a conventional superconductor, the pairing function or gap is essentially isotropic in reciprocal space.
For 
some of the pnictide superconductors,
 the Fermi surface comprises both electron pockets at the Brillouin zone boundary and 
hole pockets at the zone center.
Measurements
largely support the
notion that the pairing function 
actually
changes sign across the electron and hole pockets 
while maintaining a fully-gapped single-particle excitation spectrum,
 as predicted
  in a variety of theoretical studies\cite{basics}.
 However, phase-sensitive  measurements for
  this sign change 
remain a challenge
while other pairing channels are close by in energy. 
For the highest-$T_c$ group of iron chalcogenides, in which the 
Fermi surface contains only the electron pockets (see Box 1), similar conclusions regarding the pairing states 
have been 
reached, suggesting
 that the details of the Fermi surface are not as crucial as the effect of electron 
correlations.

The advent of orbital-selective correlations in the normal state naturally raises the question of its implications 
for the pairing amplitude, symmetry and structure in the superconducting state. 
Indeed, 
such
 orbital-selective pairing has been visualized, in spectacular fashion, via 
 tunneling measurements performed on the 
nematic FeSe (Fig.\,4) \cite{seamus}. The gap anisotropy can be understood 
 in terms of 
a dominant anisotropic
 pairing 
 amplitude 
  (with a gap that becomes very small
  at specific loci on the Fermi surface),
 once the variation of the orbital weight on the Fermi surface is taken into 
account 
 \cite{ossc}.
 We note that ARPES measurements have yet to reach a consensus on the orbital dependence
 of the pairing amplitude in FeSe.

\section{New Horizons}
\label{sec:new_horiz}	

Space restrictions inevitably prevent us from covering the whole gamut of exciting physics 
that is emerging from research into iron-based 
superconductors, such as proximity of their superconducting state to a Bose-Einstein condensate 
or the prospects for observing topological superconductivity at elevated temperatures. 
With regards the latter, we simply note that
FeSCs hint at a possible design principle whereby superconductivity can be enhanced 
by quantum critical fluctuations within a topological setting.
Rather, in
 this final section, we 
consider 
the major impact that FeSCs are having in shaping our understanding of 
other
disparate families of unconventional superconductors.

As emphasized 
in the introduction,
 the insights gained from the field of FeSCs are also guiding efforts to understand a variety 
of other superconductors. Chief among these is the guiding principle, 
so aptly demonstrated by the BaFe$_2$(As$_{1-x}$P$_x$)$_2$
series, that superconductivity is enhanced 
near a QCP (Box 2). These developments have solidified the thought that 
quantum critical fluctuations promote unconventional superconductivity, 
and have led to the notion that orbital multiplicity allows for
new types of Cooper pairing.
 In this regard, developments 
in the FeSCs have impacted too on investigations into
  venerable multi-orbital 
superconductors. While Sr$_2$RuO$_4$ has long been championed as 
a
 chiral 
spin-triplet superconductor, recent experiments have suggested that it is a 
spin-singlet superconductor with an unusual pairing function; the ideas being put forward for candidate 
pairing functions to some degree parallel those for the FeSCs.
Another multi-orbital superconductor, CeCu$_2$Si$_2$ -- the very first unconventional superconductor ever discovered --
has long been considered to be a nodeless superconductor. Recent experiments down to lower temperatures, however, 
have provided 
compelling evidence that the gap does not close anywhere on the Fermi surface, even though it is strongly
anisotropic. The leading idea to resolve this conundrum invokes 
a multi-orbital pairing state that is analogous to what has been proposed for the iron 
chalcogenides
\cite{multiorbital-sc}. 
 The
 role of multi-orbital physics is now being actively investigated in a host of other, 
more recently discovered superconductors, including
nickel-based compounds that represent
a close structural cousin to the high-$T_c$ cuprates, the exotic heavy-fermion compound UTe$_2$
as well as the V-based kagome metals.

The prominent role that nematicity plays in the FeSCs 
not only
 provides
 a boon for uniaxial strain measurements 
as a means of tuning and probing unconventional 
superconductivity, it has also encouraged the community to consider new pathways to improved superconducting performance. 
The essential argument is as follows:
though the electron-phonon interaction acts on the entire Fermi surface, it is inherently weak. 
By contrast, antiferromagnetic interactions tend to be much stronger, but only influence a restricted region in momentum space.
 In principle, nematicity offers the best of both worlds, i.e. strong interactions acting on the entire Fermi sea. 
 The challenge then is to harness this interaction either on its own or in conjunction with another pairing instability.
At the same time, the success of engineered structures based on 
monolayer
FeSe in raising $T_c$ has provided added incentive to achieve superconductivity in the purely two-dimensional 
limit, a frontier that has enjoyed remarkable recent success
with the discovery of superconductivity in monolayer WTe$_2$ and twisted bilayer graphene to name but two.
One might even argue that this has motivated other
recent global efforts in realizing superconductivity under extreme conditions, such as ultra-high pressures for 
room-temperature superconductivity in the hydrides, 
or
 the realms of ultra-low temperatures for superconductivity driven by quantum criticality 
in YbRh$_2$Si$_2$.
 
Given this rich vein of commonality, time seems ripe for the development 
of a conceptual framework that unifies our 
understanding of these disparate material families of unconventional superconductors, 
with the precocious pnictides firmly at the helm.
Ultimately, we would like to know how to achieve superconductivity with even higher transition temperatures
at ambient conditions.
Is there a design principle for boosting superconductivity? Our considerations in this article suggest the need for
two central ingredients 
that cooperate with each other. One is for the entire Fermi surface to participate in promoting superconductivity. 
The other is to maximize the strength of the effective interactions that drive the superconductivity.  
The band of unconventional superconductors to which the FeSCs belong
have the characteristic property that all electronic states are on the verge of localization
and thus experience strong coupling. Compared to good metals, they also host stronger interactions that favor superconductivity.
  We
envision
   that 
tuning the balance between interaction strength and localization is
 precisely the 
tool
 required to optimize
  superconductivity.

\vspace{0.5cm} \noindent{\bf Acknowledgments}
We thank Anna B\"ohmer, Lei Chen, Amalia Coldea, Pengcheng Dai, 
Seamus Davis, Ian Fisher, Frederic Hardy, Christoph Meingast, 
Matthew D. Watson, Ming Yi, Rong Yu,  and Jian-Xin Zhu
for their input. The work has been supported in part by the DOE BES Award 
No. DE-SC0018197
and the European Research Council 
(ERC) under the European Union's Horizon 2020 research and innovation programme (Grant Agreement No. 835279-Catch-22).

\newpage



\clearpage
\newpage


\begin{figure}[t!]
\centering

\includegraphics*[width=0.56\textwidth]{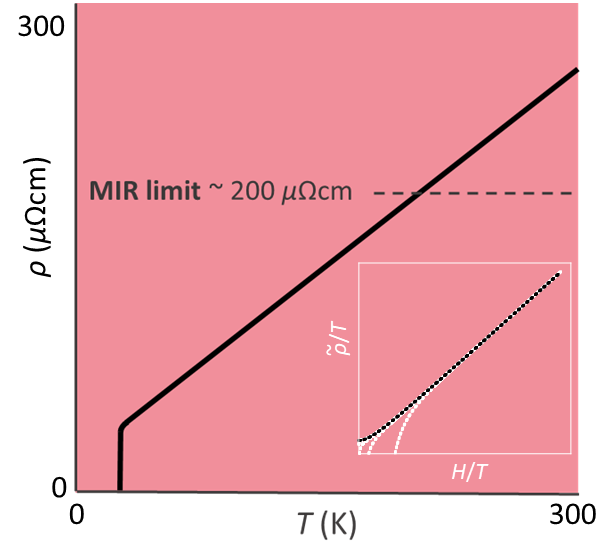}
\vspace{0.2cm}

\caption{\label{fig:Tlinear-MIR} 
{\bf The allure of linearity.}
While the temperature dependence of the resistivity $\rho(T)$ of a typical metal is linear at intermediate temperatures 
(e.g. between 100 and 400 K), it weakens and eventually vanishes at the extremes. 
At low-$T$, $\rho(T)$ levels off due to the freezing out of the phonons, while at high-$T$, $\rho(T)$ eventually saturates 
once the electron mean-free-path $\ell$ shrinks to the size of the lattice spacing $a$ or the
de Broglie wavelength ${2 \pi/k_F}$.
a threshold known as the Mott-Ioffe-Regel (MIR) limit. 
In a similar vein, the
magnetoresistance of an ordinary metal
changes quadratically with magnetic field strength $H$ at low $H$, is quasi-linear over a narrow field 
range before tending towards 
another
 saturation value 
 (albeit unrelated
  to the MIR limit). By contrast, the in-plane
resistivity of the iron pnictide 
BaFe$_2$(As$_{0.7}$P$_{0.3}$)$_2$
 exhibits  linearity
 over an extremely broad 
  range of temperatures (main panel, adapted from Kasahara \emph{et al.} \cite{Hayes})
and magnetic field (inset, adapted from Hayes \emph{et al.} \cite{Hayes}).
At high temperatures,
 the magnitude of the resistivity exceeds the MIR
 limit,
signifying that electron correlations are strong \cite{hussey04}.
The magnetoresistance also
exhibits a unique form of scaling
whereby plots of $\tilde{\rho}/T$ vs. $H/T$ collapse 
onto a single line. 
(The downturns at low field are due to the onset of
superconductivity.)
Here $\tilde{\rho} =  \rho (H, T) - \rho (0, 0)$.
This contrasts starkly with the so-called Kohler scaling found in conventional metals where both $\rho(0,0)$ and $T$ 
are replaced by $\rho(0,T)$. 
Such simple linear scaling hints at a highly anomalous metallic state in which impurity scattering, 
clearly evident in the zero-field resistivity, plays no role in the evolution of the resistivity in an applied magnetic field.
}
\end{figure}
\clearpage
\newpage

\begin{figure}[t!]
\centering

\includegraphics*[width=0.9\textwidth]{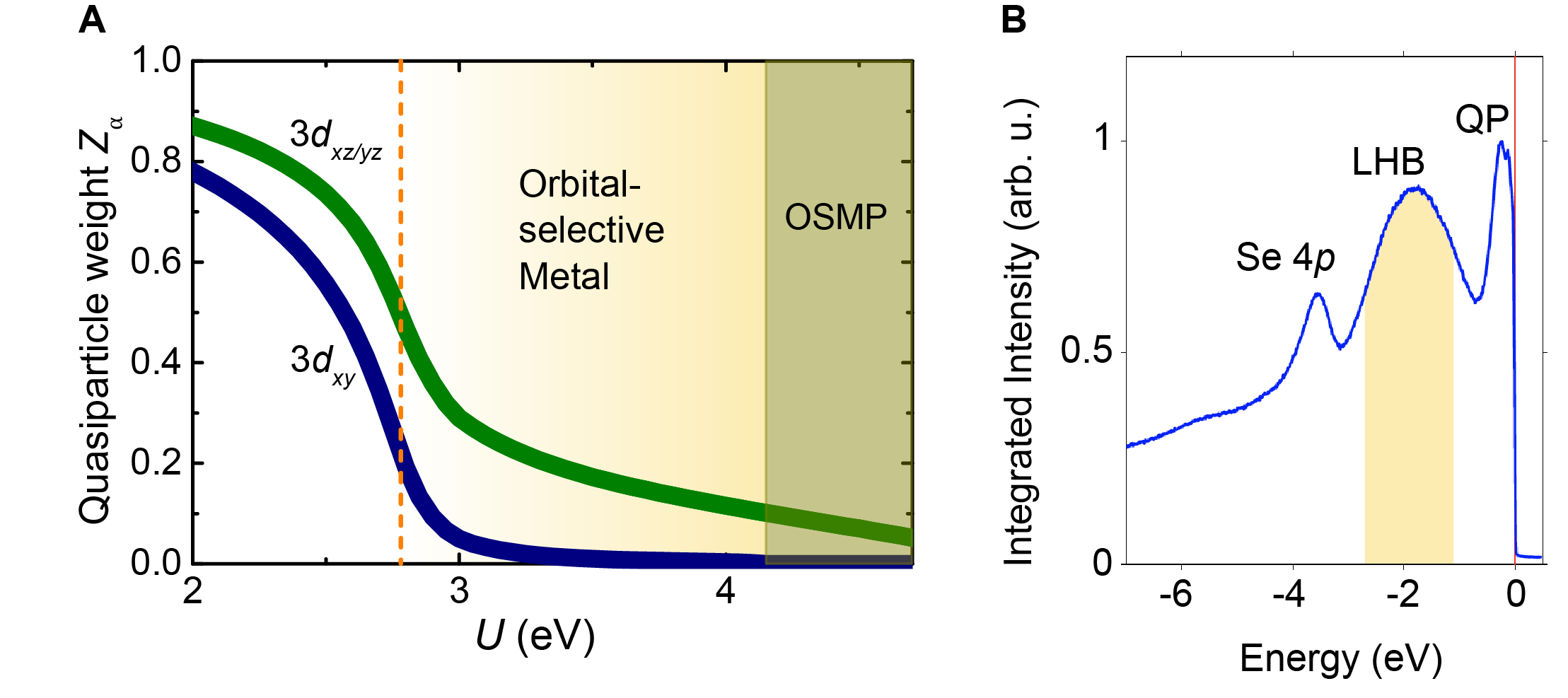}
\vspace{0.2cm}

\caption{\label{fig:os-corr} 
{\bf Correlations are (orbital-)selective.}
In a simple metal like Aluminium, conduction electrons are essentially interaction-free. 
They readily move across the entire system through a wave-like propagation. In FeSCs, the electrostatic 
repulsion between the
electrons serves as the primary driver for their superconductivity (see Fig.\,\ref{fig:os-sc}). The electron-electron
interactions also cause correlations in the electron motion. The correlated electrons act as if only a fraction, $Z$,
is freely propagating. Moreover, in FeSCs, this
 ``quasiparticle weight" is orbital-dependent. Shown in ({\bf A}) is $Z_{\alpha}$, for the orbitals
$\alpha=3d_{xy}$ and $3d_{xz/yz}$  as a function of the onsite intra-orbital Coulomb repulsion $U$ (for an inter-orbital
Hund's coupling $J_H=0.2U$). Over an extended interaction range, the quasiparticle weight of the electrons in
the $3d_{xy}$ orbitals is much reduced compared with those of the other orbitals. This regime of orbital-selective metal 
is anchored by an orbital-selective  Mott phase (OSMP), in which the $3d_{xy}$ electrons are fully localized while the other 
orbitals still have a nonzero quasiparticle  (metallic) weight. 
Proximity to the OSMP means that the charge degrees of freedom
are on the verge of being localized. Evidence for the latter has emerged from ARPES measurements ({\bf B}).
In addition to noninteracting $4p$ states from the Se atoms and a quasiparticle (QP) peak near the Fermi energy 
(marked as $0$ in panel B), we find
 a ``lower Hubbard band" 
 peak (LHB)
  associated with the excitations between 
 atomic energy levels
 split by this intra-orbital Coulomb repulsion.
Adapted from
Yu \emph{et al.} \cite{theory-osm}
 and Watson \emph{et al.} \cite{watson-lhb}.
}
\end{figure}
\clearpage
\newpage

\begin{figure}[t!]
\centering

\includegraphics*[width=0.58\textwidth]{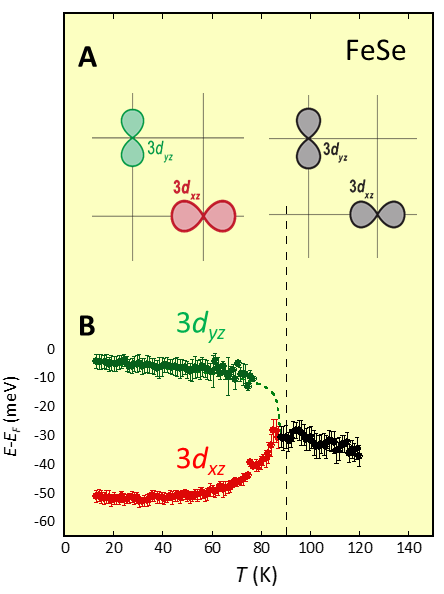}
\vspace{0.2cm}

\caption{\label{fig:nematic}  {\bf 
Nematicity in FeSe.}
{\bf A)} 
Schematic of the breaking of orbital degeneracy as proxy for
the
electronic nematic transition across $T_S$ (marked by the vertical dashed line).
The constituent ions of a crystal are arranged according to a geometric lattice, providing a periodic potential 
for the mobile electrons. The crystalline symmetry of FeSe dictates that its $3d_{xz}$ and $3d_{yz}$ orbitals can be 
transformed into each other by a 90 degree rotation within its tetragonal plane. This is indeed the case for
 $T > T_S$.
 Upon cooling below $T_S$, however, 
 this symmetry is broken through the formation of an electronic nematic. The symmetry breaking 
 is manifest in terms of an inequivalence between the electronic bands associated with the $3d_{xz}$ and $3d_{yz}$ orbitals,
 which are now marked by different colors.
{\bf B)} The $3d_{xz}$ and $3d_{yz}$ bands obtained from 
ARPES measurements on detwinned 
FeSe, showing degenerate bands that are split as temperature is lowered through
the nematic transition at $T_S$
 (adapted from Yi \emph{et al.} \cite{Fig3}). 
}
\end{figure}
\clearpage
\newpage

\begin{figure}[t!]
\centering

\includegraphics*[width=0.6\textwidth]{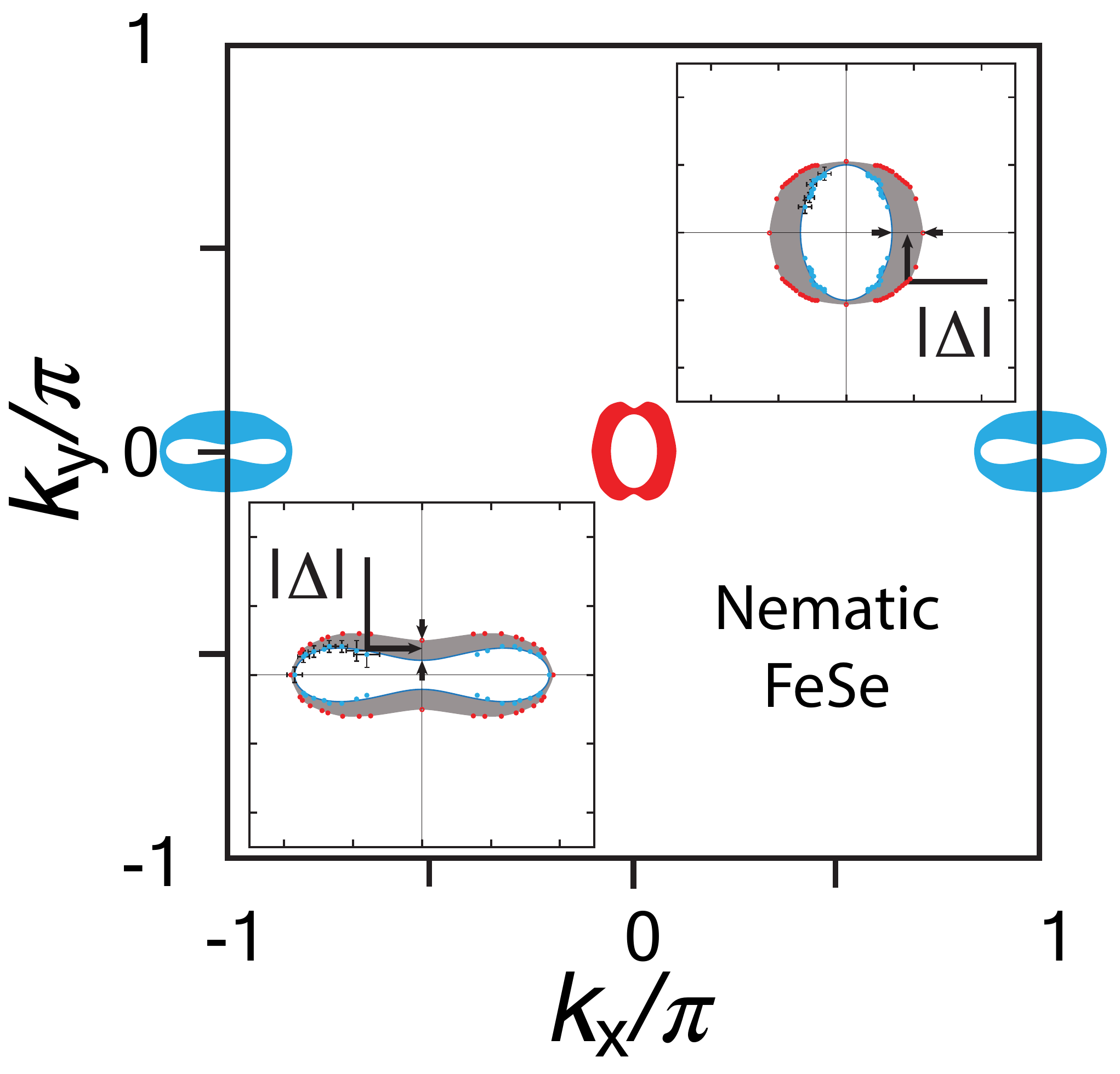}
\vspace{0.2cm}

\caption{\label{fig:os-sc} 
{\bf Superconducting pairing from (orbital-)selective correlations.}
Variation of the 
superconducting (SC)
gap magnitude on the hole Fermi pocket at the center $\Gamma$ (red, 
and top inset) and on the electron Fermi pocket at the corner $M_x$ (blue, and bottom inset) 
of the Brillouin zone
in FeSe, as determined from 
scanning tunneling microscopy measurements
 (adapted from
Sprau \emph{et al.} \cite{seamus}).
In a conventional phonon-mediated superconductor, the SC
gap is essentially isotropic, i.e. of the same magnitude, over the entire Fermi surface. 
In FeSe, on the other hand, it is clear that the SC gap almost vanishes for specific momenta. 
Such strong anisotropy not only hints at a non-phonon pairing mechanism, 
it also illustrates how the different orbitals that contribute electronic states 
at the Fermi level can have very different pairing strengths.
}
\end{figure}
\clearpage
\newpage

\begin{InfoBox}[h]
\caption{{\bf Iron-based superconductors -- some basics}\label{box_basics}} 
\fbox{
\begin{minipage}{0.98\textwidth}\raggedright
The various FeSC family 
members, together
 with
their maximal $T_c$ values
are summarized in panel {\bf A}.
The highest $T_c$ appears in monolayer FeSe deposited on a SrTiO$_3$ substrate (FeSe/STO).
The accepted value of the $T_c$  record, which is based on the onset of the Meissner effect, 
is $65$ K, though transport evidence for $T_c$ above $100$ K has also been reported.
 
All FeSCs have the same structural motif, a single layer of FeSe/FeTe or FeAs/FeP.  
This is illustrated by the structure of the bulk FeSe, which corresponds to a direct stacking 
of FeSe layers (panel {\bf B}). The primitive unit cell of a single FeSe/FeAs layer is shown 
in panel {\bf C}, with two Fe-ions from an Fe-layer, one Se/As ion located above the Fe-layer 
and one below. When spin-orbit coupling is neglected (see Sec.\,II), the Brillouin zone can be unfolded to 
a square in reciprocal (${\bf k}$) space, shown in panel {\bf D}. The unfolded Brillouin zone
corresponds to the square lattice of Fe-ions illustrated in panel {\bf C}. Strictly speaking, 
one needs to use the physical two-Fe unit cell and its associated Brillouin zone, which is 
half of what is shown in panel {\bf D}. While it is rigorous, this notation is also somewhat 
cumbersome.
Usually, it is more convenient to adopt the single-Fe unit cell and its associated 
Brillouin zone. 
Correspondingly, microscopic theoretical studies typically involve 
multi-orbital models on a square lattice with both onsite 
Hubbard (direct Coulombic) and Hund's (spin-exchange)
interactions.

The electronic states near the Fermi energy are dominated by the $3d$ orbitals of the Fe-ions. 
Thus, for most purposes, theoretical models of FeSC retain only the $3d$ states, with the $p$-orbitals 
of Se/As-ions (or, their variants, Te/P-ions) projected out. The multiplicity of the $3d$ orbitals 
near the Fermi energy is reflected in the multiple Fermi sheets. In most FeSCs, there are hole 
Fermi pockets near the center (or $\Gamma$) of the one-Fe Brillouin zone and electron Fermi pockets 
at the edge ($M_x$ and $M_y$) of the one-Fe Brillouin zone. 
The hole and electron Fermi pockets have roughly the same size, 
which enhances the phase
space for inter-pocket electron interactions; they are called ``nested".
In the single-layer FeSe/STO, there 
are only electron Fermi pockets near $M_x$ and $M_y$, as shown in panel {\bf D}. The same is 
true in most members of the bulk iron chalcogenides with relatively high $T_c$, including the 
alkaline iron selenides, where $T_c$ reaches about $30$ K, and the Li-intercalated iron selenides 
(Li,Fe)OHFeSe, whose $T_c$ exceeds $40$ K. These members are marked in blue in panel {\bf A}.

\end{minipage}
}
\end{InfoBox}
\clearpage
\newpage


\fbox{
\begin{minipage}{0.98\textwidth}

\centerline{
\includegraphics[height=0.68\textwidth]{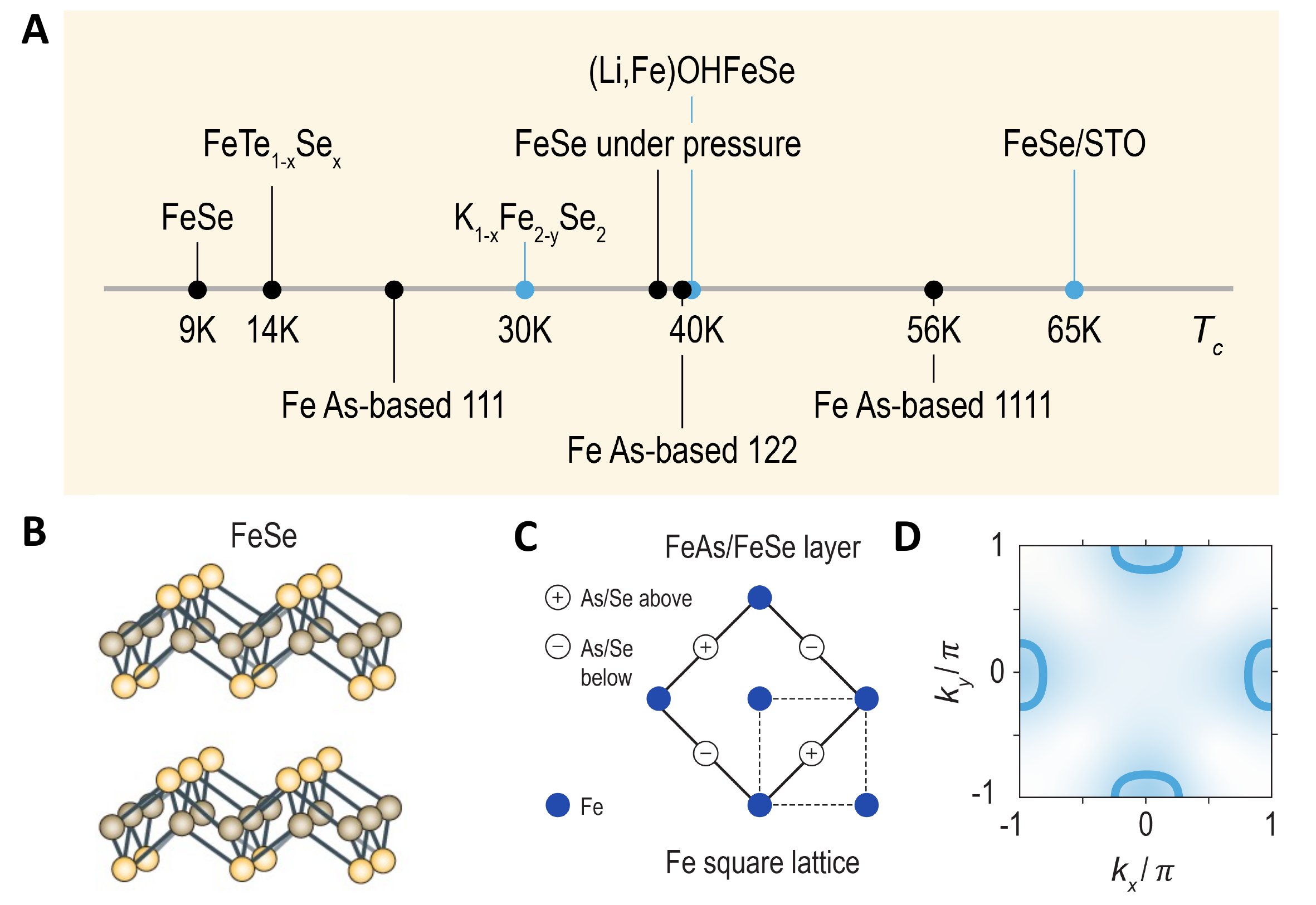}
}


\end{minipage}
}
\clearpage
\newpage


\begin{InfoBox}[h]
\vspace{-0.8cm}

\caption{{\bf Quantum criticality in iron pnictides -- where all the players come together}\label{box_players}} 
\fbox{
\begin{minipage}{0.98\textwidth}\raggedright
Quantum criticality has been extensively studied in the iron pnictides. An early theoretical analysis 
by Dai \emph{et al.} \cite{qcp-FeAs} 
 led to the proposal
that iso-electronic P-for-As substitution 
yields a QCP at 
a concentration $x=x_c$ (panel {\bf A}). Here, the antiferromagnetic (AF) and nematic (nem) orders
vanish concurrently; electronic nematicity describes the development of 
orientational-symmetry breaking in the electronic phase, and 
can be pictured by analogy with the anisotropy seen in
cosmic microwave background.
As a result, both the magnetic and nematic degrees of freedom play a central role 
in creating critical quantum fluctuations (panel {\bf A}) and in causing the effective carrier mass 
to diverge (panel {\bf B}). The concurrent quantum phase transitions have now been observed in both 
P-substituted CeFeAsO and BaFe$_2$As$_2$.

In the BaFe$_2$(As$_{1-x}$P$_x$)$_2$ series, quantum criticality has been demonstrated in multiple ways. 
These include the 
extended $T$-linearity
of the
 electrical resistivity
 and an
 effective 
 mass that
  diverges
 as the QCP is approached from the paramagnetic side ($x>x_c$)
\cite{qcp-FeAs}.
In 
the vicinity of the QCP,
the superconducting $T_c$ is maximized. Thus, BaFe$_2$(As$_{1-x}$P$_x$)$_2$ provides a textbook example 
of 
 superconductivity 
driven by the same critical quantum fluctuations that are responsible for the emergence of 
anomalous transport properties.

\end{minipage}
}
\par
\end{InfoBox}

\clearpage
\newpage


\fbox{
\begin{minipage}{0.98\textwidth}

\centerline{
\includegraphics[height=0.88\textwidth]{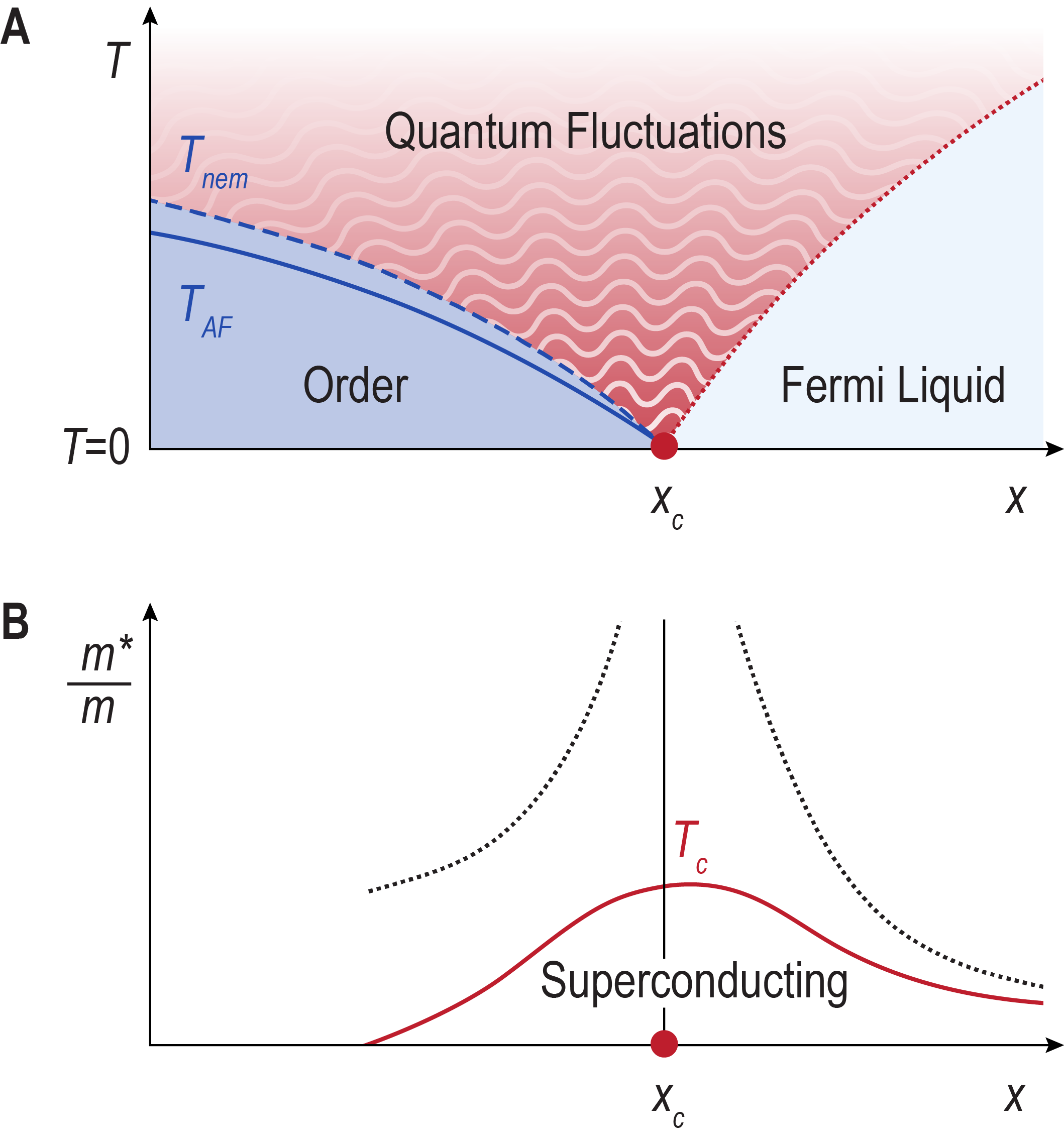}
}


\end{minipage}
}
\newpage

\end{document}